\documentclass[aps, prl, preprintnumbers, showpacs, twocolumn, superscriptaddress,letterpaper,nofootinbib]{revtex4}

\usepackage{graphicx}
\usepackage{dcolumn}
\usepackage{amssymb}
\usepackage{amsmath}
\usepackage{amsfonts}
\usepackage{dcolumn}
\usepackage{natbib}
\usepackage{color}

\def\apj{Astrophys. J.}
\def\apjl{Astrophys. J. Lett.}

\def\aap{Astron. Astrophys. }

\def\mnras{Mon. Not. Roy. Astron. Soc. }

\def\prd{Phys. Rev. D.}

\def\cqg{Class. Quant. Grav.}

\def\aap{Astron. Astrophys. }

\begin{document}

\title{Dynamics and Gravitational Wave Signature of Collapsar Formation}

\author{C.~D.~Ott}
\email{cott@tapir.caltech.edu}
\affiliation{TAPIR, Mail Code 350-17, California Institute of Technology,
  Pasadena, CA, 91125, USA}

\author{C. Reisswig}
\affiliation{TAPIR, Mail Code 350-17, California Institute of Technology,
  Pasadena, CA, 91125, USA}

\author{E. Schnetter}
\affiliation{Center for Computation \& Technology, Louisiana State
  University, Baton Rouge, LA 70803, USA}
\affiliation{Department of Physics \& Astronomy,
  Louisiana State University, Baton Rouge, LA 70803, USA}

\author{E.~O'Connor}
\affiliation{TAPIR, Mail Code 350-17, California Institute of Technology,
  Pasadena, CA, 91125, USA}

\author{U. Sperhake}
\affiliation{Institut de Ci\`encies de l'Espai (CSIC-IEEC), Facultat de Ci\`encies, Campus UAB, E-08193 Bellaterra, Spain}
\affiliation{TAPIR, Mail Code 350-17, California Institute of Technology,
  Pasadena, CA, 91125, USA}

\author{F. L\"offler}
\affiliation{Center for Computation \& Technology, Louisiana State
  University, Baton Rouge, LA 70803, USA}

\author{P. Diener}
\affiliation{Center for Computation \& Technology, Louisiana State
  University, Baton Rouge, LA 70803, USA}
\affiliation{Department of Physics \& Astronomy,
  Louisiana State University, Baton Rouge, LA 70803, USA}

\author{E. Abdikamalov}
\affiliation{Center for Computation \& Technology, Louisiana State
  University, Baton Rouge, LA 70803, USA}

\author{I. Hawke} \affiliation{School of Mathematics, University of
  Southampton, Southampton, UK}

\author{A. Burrows} \affiliation{Department of Astrophysical Sciences,
  Princeton University, 4 Ivy Lane, Princeton, NJ 08544, USA}


\begin{abstract}
  We perform $3+1$ general relativistic simulations of rotating core
  collapse in the context of the collapsar model for long gamma-ray
  bursts. We employ a realistic progenitor, rotation based on results
  of stellar evolution calculations, and a simplified equation of
  state. Our simulations track self-consistently collapse, bounce, the
  postbounce phase, black hole formation, and the subsequent early
  hyperaccretion phase. We extract gravitational waves from the
  spacetime curvature and identify a unique gravitational wave
  signature associated with the early phase of collapsar formation.
\end{abstract}

\maketitle


There is strong observational evidence linking long gamma-ray bursts
(LGRBs) with the death of massive stars in core collapse~(e.g.,
\cite{wb:06}). It appears likely that LGRBs are made in metal-poor
progenitors with degenerate iron cores. These may be ordinary massive
stars turned into Wolf-Rayet objects by mass loss or binary
interactions~\cite{fryer:99b,fryer:05} or, perhaps, peculiar,
fully-mixed stars~\cite{woosley:06,yoon:06}.  Both could result in a
type-Ibc core-collapse supernova (CCSN) harboring
a LGRB central engine. The nature of the latter and the details of the
CCSN-LGRB relationship are uncertain. Viable engine {settings} 
all require rapid progenitor rotation and include the proto\-magnetar
model~(e.g.,~\cite{bucciantini:09}) and the collapsar
scenario~\cite{woosley:93,macfadyen:99}. In the latter, the CCSN
fails and a black hole (BH) with an accretion disk forms
or a weak explosion occurs leading to fallback and
BH/disk formation.

In this Letter, we address, for the first time in $3+1$ general
relativity (GR), the formation of spinning BHs in failing CCSNe in the
context of the collapsar scenario of LGRBs.  
Our full GR method allows us to self-consistently follow 
core collapse, bounce, postbounce evolution, protoneutron star
(PNS) collapse, BH formation, and the subsequent early hyperaccretion
phase. For the first time, we extract the gravitational
wave (GW) signature of a failing CCSN that evolves into a collapsar and
track the properties of the nascent BH with the dynamical horizon
formalism~\cite{schnetter:06}.

Previous work on BH formation in CCSN/LGRB progenitors was limited to
spherical symmetry \cite{sumiyoshi:08,fischer:09a,oconnor:10b} and, due
to gauge choices, simulations could not be continued beyond
BH formation. Multi-D studies either considered isolated NS
collapse (e.g., \cite{baiotti:04}) or BH formation in very massive 
polytropes \cite{sekiguchi:05,liu:07}. Recently, Sekiguchi \&
Shibata \cite{sekiguchi:10c} carried out the first axisymmetric (2D)
GR simulation that continued beyond BH formation in a hot polytrope,
but did not extract the GW signal.


\emph{Method.}  We employ the Zelmani $3+1$ GR core collapse
simulation package \cite{reisswig:10ccwave} which is based on
the Cactus framework 
and the Carpet adaptive mesh
refinement (AMR) driver \cite{Schnetter-etal-03b}, and uses the
open-source EinsteinToolkit 
for GR curvature
(via \cite{ES-Brown2007b}) and hydrodynamics evolution (via an updated
variant of \cite{baiotti:04}).
We extract GWs directly from the spacetime fields using the fully
gauge-invariant Cauchy-Characteristic Extraction method of
\cite{reisswig:10ccwave,reisswig:10a}.
The simulations are performed in an octant of the Cartesian 3D cube
with periodic boundaries on two of the inner faces of the octant and
reflection symmetry about the equatorial plane. This limits 3D
structure to even $\ell$ and $m$ that are multiples of $4$.  We use 11
levels of AMR, adding levels during collapse and postbounce evolution
when needed.
In our baseline resolution (BR), the finest resolution is
$\sim$$370\,\mathrm{m}$ and $\sim$$92~\mathrm{m}$ at bounce and BH
formation, respectively. We also perform calculations with $20\%$
higher/lower (HR/LR) resolutions and check stability and consistency
by monitoring the ADM constraints. They show
2$^\mathrm{nd}$-order convergence up to bounce and
$1^\mathrm{st}$-order afterwards. After BH formation, convergence is
reduced near the singularity, but the simulations remain consistent
and stable. ADM mass and angular momentum are conserved to $\lesssim
3\%$ in BR runs. All runs are carried out past BH formation, but only
LR runs are continued to tens of ms after BH formation.

We employ a hybrid polytropic--$\Gamma$-law equation of state
(EOS; e.g., \cite{janka:93}). It smoothly matches a polytrope described
by $\Gamma_1 \approx 4/3$ at subnuclear densities with one described
by $\Gamma_2 > \Gamma_1$ at supernuclear densities, allowing to
capture the stiffening of the nuclear EOS. A $\Gamma$-law component
(described by $\Gamma_\mathrm{th}$) accounts for thermal pressure
contributions due to shock heating.  We set $\Gamma_1 = 1.31$ in the
collapse phase
and choose a rather soft
supernuclear EOS by setting $\Gamma_2 = 2.4$.
This results in a maximum non-spinning PNS gravitational mass of $\sim
1.7\,M_\odot$, which provides for rapid BH formation, but is below the
empirical NS mass limit~\cite{demorest:10}.
We choose $\Gamma_\mathrm{th} = 4/3$ for the postshock flow
whose effective $\Gamma$ is reduced by the dissociation of
Fe-group nuclei.
Neutrino heating (unlikely to be dynamically relevant in this
scenario) is neglected, but we account for postbounce neutrino
cooling of the outer PNS and the postshock region via the cooling
function given in \cite{murphy:09}.

\begin{figure}[t]
\includegraphics[width=7.8cm]{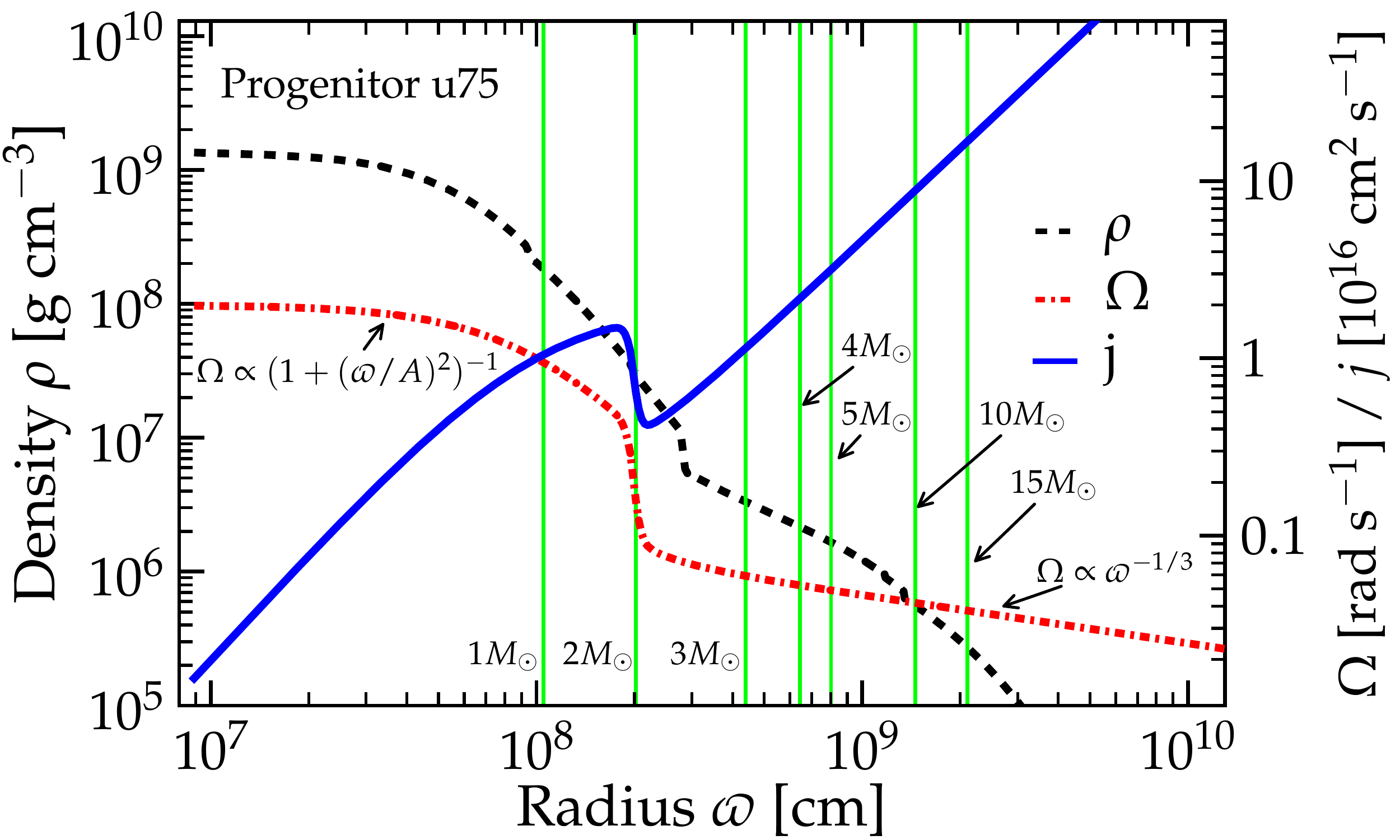}
\caption{Progenitor model u75. Left ordinate:
  Radial density distribution. Right ordinate: Angular velocity (red
  dash-dotted curve) and specific angular momentum (blue curve) as a
  function of cylindrical radius $\varpi$ as given by our rotation
  law, reproducing features seen in the rotating progenitors of
  \cite{woosley:06}. Vertical lines mark the enclosed mass.} \label{fig:prog}
\end{figure}

\emph{Initial Conditions.}  We use the 75-$M_\odot$,  $10^{-4}$-solar
metallicity model u75 of \cite{whw:02} whose compact core favors early
BH formation \cite{oconnor:10b}. u75 could be a viable GRB progenitor
if mass transfer to a binary companion removed its H/He envelopes. We
map u75's inner $\sim$$5700\,\mathrm{km}$ (enclosed mass
$\sim$$4.5\,M_\odot$) onto our 3D grid and impose constant rotation on
cylindrical shells with radius $\varpi$ via a rotation law motivated
by the GRB progenitors of \cite{woosley:06}: The inner iron core is in
near uniform rotation and $\Omega$ drops $\propto \varpi^{-2}$ further
out.
Close to the edge of the iron core, $\Omega$ drops by a factor of
order unity, then continues to decrease $\propto r^{-\zeta}$, with
$0 < \zeta < 2$, leading to a radial increase in the specific angular
momentum $j$, endowing mantle
material with sufficient spin to form a disk at small radii. The
functional form is
$ \Omega(\varpi) = (1-\lambda(\varpi)) 
\Omega_0 (1+(\varpi/A)^2)^{-1} 
+ \lambda(\varpi) \xi \Omega_0 (1 + (\varpi_\mathrm{t}/A)^2)^{-1}
(1+(\mathrm{max}(0,\varpi - \varpi_\mathrm{t})/A))^{-\zeta}$.  
Here,
$\lambda(\varpi) = (1 + \tanh((\varpi-\varpi_\mathrm{t})/\delta\varpi))/2$.  
We set $A =
1000\,\mathrm{km}$, $\varpi_t = 1950\,\mathrm{km}$, $\xi = 1/3$, and
$\delta \varpi = 100\,\mathrm{km}$. $\Omega_0$ is the central angular
velocity that we vary from $0$ to $2\,\mathrm{rad\,s}^{-1}$.  
Fig.~\ref{fig:prog} depicts u75's density profile along with
$\Omega(\varpi)$ and $j(\varpi)$ for the $\Omega_0 =
2\,\mathrm{rad\,s}^{-1}$ case.  Model names, parameters, and key
results are given in Table~\ref{tab:models}.


\begin{figure}[t]
\includegraphics[width=8.5cm]{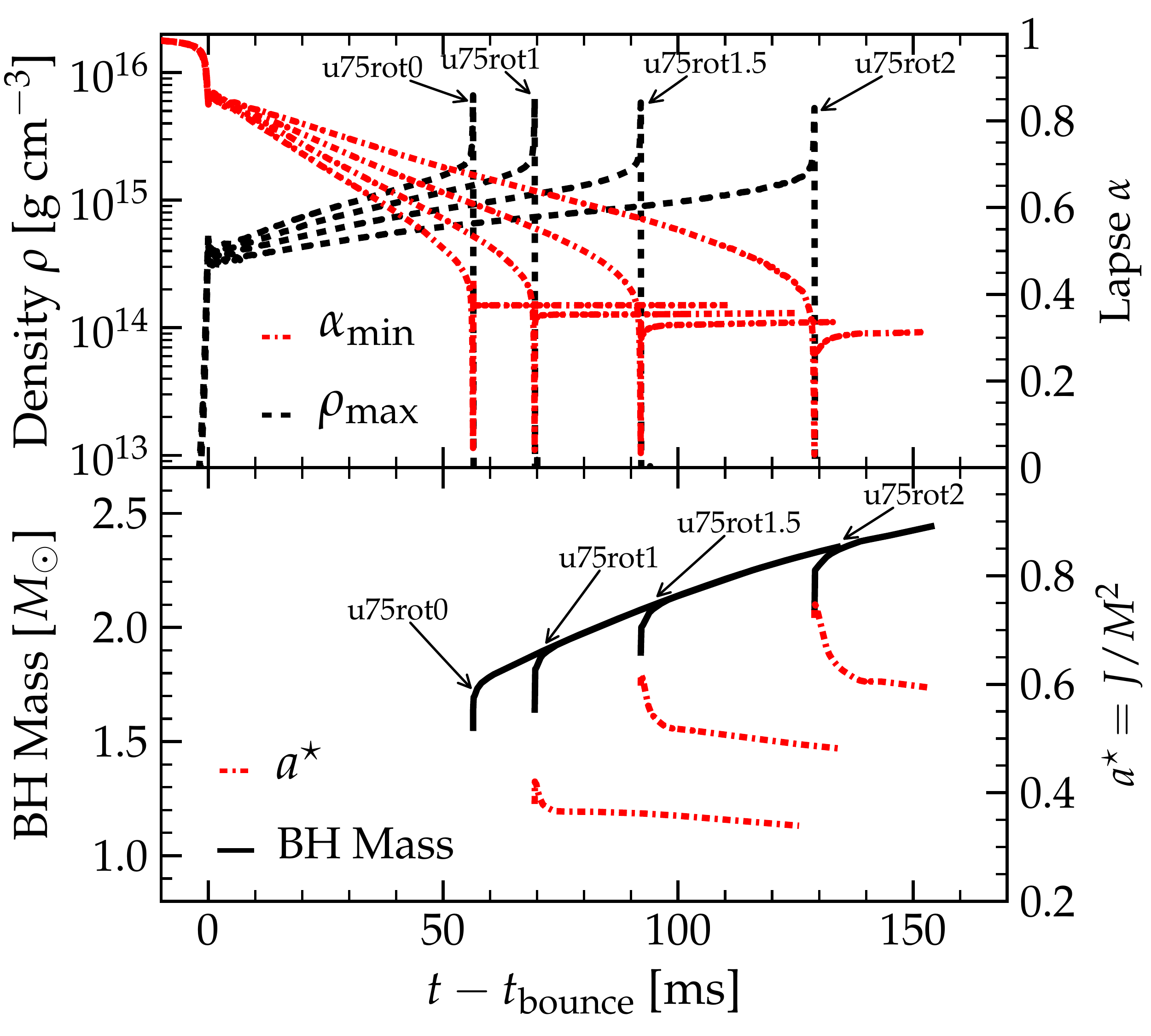}
\caption{{Top}: Maximum density $\rho_\mathrm{max}$ and central
ADM lapse function $\alpha_\mathrm{min}$ as a function of postbounce
time in all models. After horizon formation, the region interior to it
is excluded from min/max finding.  {Bottom}: BH mass and
dimensionless spin $a^\star$ as a function of postbounce time. All
models follow the same accretion history once a BH forms and settles
down.  The data shown in this figure are from the LR runs since these
were carried out longest after BH formation. 
\label{fig:evol}}
\end{figure}

\begin{table}
  \caption{Model summary. $\Omega_0$ is the initial central angular
    velocity. $t_\mathrm{BH}$ is the time after bounce to BH
    formation. $M_\mathrm{max}$ is the mass of the PNS at that
    time. $a^\star_i$ and $a^\star_e$ are the dimensionless BH spin
    shortly after BH formation and when the simulation is stopped,
    respectively.  $E_\mathrm{GW}$ is the emitted GW energy and $f_c$
    is the characteristic GW frequency~\cite{thorne:87} in
    aLIGO.\label{tab:models}}
\begin{ruledtabular}
\begin{tabular}{lccccccc}
Model
  &$\Omega_0\,$
  &$t_\mathrm{BH}$
  &$M_\mathrm{max}\,$
  &$a^{\star}_i$
  &$a^\star_e$ 
  &$E_\mathrm{GW}$
  &$f_c$\\
  &$[\mathrm{rad}\,\mathrm{s}^{-1}]$
  &$[\mathrm{ms}]$
  &$[M_\odot]$
  & 
  & 
  &$[10^{-7} M_\odot c^2]$ 
  &$[\mathrm{Hz}]$ \\
u75rot0  & 0.0&\phantom{1}56.4&1.69&--&--&$0.06$&591\\
u75rot1  & 1.0&\phantom{1}68.8&1.82&0.42&0.33&$1.1\phantom{6}$&838\\
u75rot1.5& 1.5&\phantom{1}92.1&2.00&0.62&0.48&$2.3\phantom{6}$&848\\
u75rot2  & 2.0&129&2.25&0.75&0.59&$3.4\phantom{6}$&807\\
\end{tabular}
\end{ruledtabular}
\end{table}


\emph{Dynamics.}---%
The homologous collapse of the inner core to nuclear densities
proceeds as in the standard CCSN case. For the initial inner core
rotation rates considered here, centrifugal effects are negligible in
the prebounce phase and all models reach core bounce after
$\sim$$114\,\mathrm{ms}$ of collapse. A hydrodynamic bounce shock is
launched, but, due to neutrino cooling and the low $\Gamma$ in the
postshock region, quickly (within milliseconds) succumbs to the ram
pressure of the outer core, which is accreting at a rate
of initially tens of $M_\odot\, \mathrm{s}^{-1}$.
The shock stalls at only $\lesssim$$50\,\mathrm{km}$ and gradually
retracts in all models.
In the top panel of Fig.~\ref{fig:evol}, we plot the 
maximum rest mass density $\rho_\mathrm{max}(t)$ that
rapidly increases as accreted material settles onto the outer PNS
core. 
The slope of $\rho_\mathrm{max}$
is steepest in the nonrotating model whose PNS becomes unstable 
earliest. In rotating models, centrifugal effects lead to an oblate and less
compact PNS that contracts more slowly and is stable to larger mass
(cf.~\cite{oconnor:10b,baumgarte:00} and Table~\ref{tab:models}).
The time to BH formation and the
maximum PNS mass increase roughly with $\Omega_0^2$. 

\begin{figure}
  \includegraphics[width=1.\linewidth]{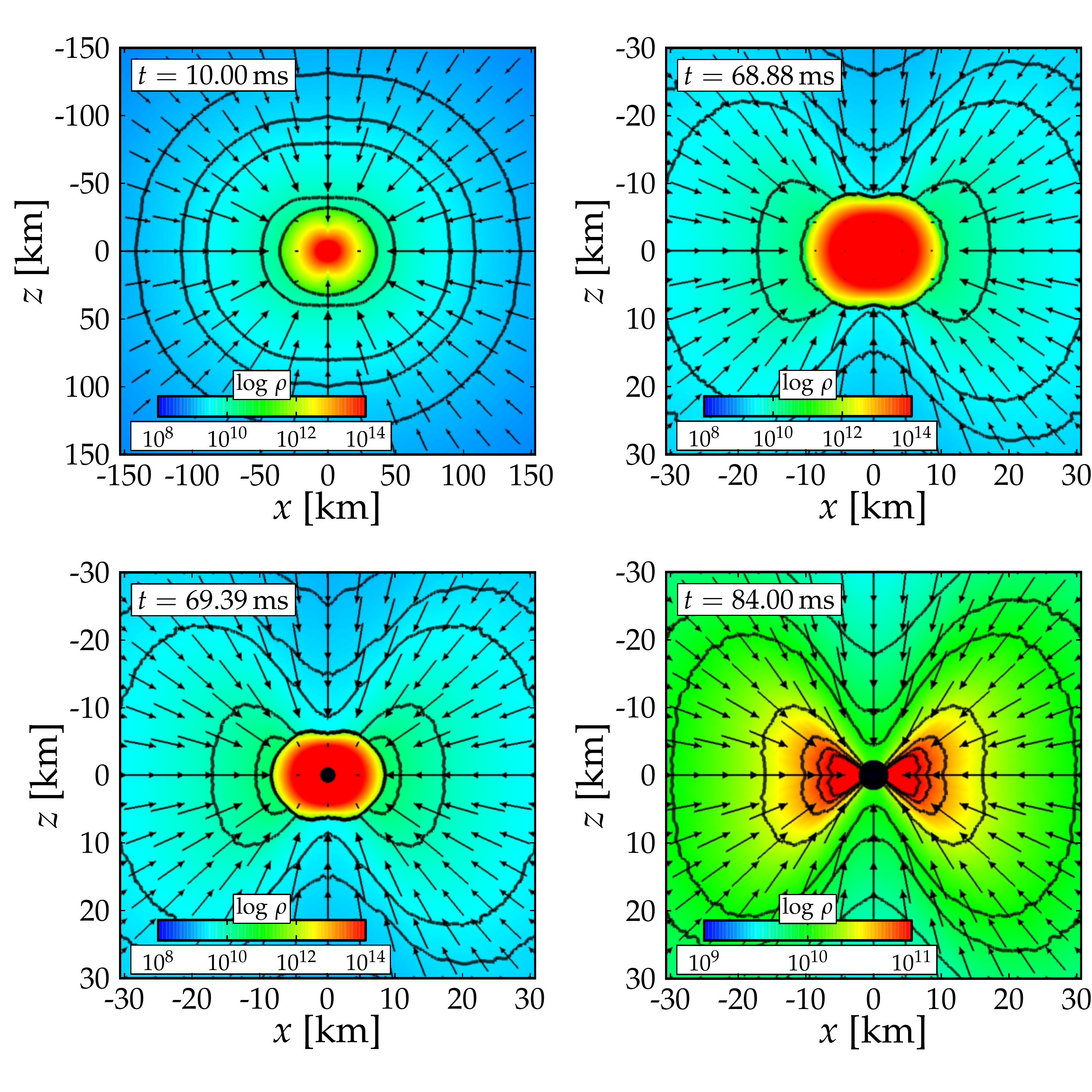}
  \vspace{-0.5cm}
  \caption{Snapshots of the meridional density distribution with
    superposed velocity vectors in model u75rot1 taken at various
    times.  The top left panel (note its special spatial range) shows
    a snapshot from $10\,\mathrm{ms}$ after bounce. The top right and
    bottom left panels show the point of PNS instability and the time
    at which the AH first appears, respectively.  The bottom right
    panel, generated with a separate color range, shows the
    hyperaccreting BH at $\sim 15\,\mathrm{ms}$ after its
    formation. All colormaps have density isocontours superposed at
    densities (from outer to inner) of $\rho=(0.1,
    0.25,0.5,0.75,1.0,2.5,5.0)\,\times 10^{10}\,\rm{g}\,
    \rm{cm}^{-3}$.}
  \label{fig:rot2}
\end{figure}

Once dynamical PNS collapse sets in, an apparent
horizon 
(AH) appears within $\sim$$1\,\mathrm{ms}$ and quickly engulfs the entire
PNS.
With the PNS and pressure support removed, postshock material and the
shock itself immediately subside into the nascent BH. 
The bottom panel of
Fig.~\ref{fig:evol} shows the evolution of BH mass and dimensionless
spin $a^\star$ in all models. The former jumps up as the AH swallows
the PNS and postshock region, then increases at the rate of accretion
set by progenitor structure and is largely unaffected by rotation at
early times.  The dimensionless spin reaches a local maximum when the
BH has swallowed the PNS core, then rapidly decreases as surrounding
lower-$j$ material plunges into the BH. This is a consequence of the
drop of $j$ at a mass coordinate close to the initial BH mass
(cf.~Fig.~\ref{fig:prog}). Table~\ref{tab:models} summarizes for all
models the values of $a^\star$ at its peak and at the time we stop the
LR run.

In Fig.~\ref{fig:rot2}, we plot colormaps of the density in the
meridional plane of the spinning model u75rot1 taken at various
postbounce times.  The rotational flattening of the PNS is significant
and so is the centrifugal double-lobed structure of the
post-BH-formation hyperaccretion flow. The latter is unshocked and far
sub-Keplerian with inflow speeds of up to $0.5 c$ near the
horizon. The flow will be shocked again only when material with
sufficiently high specific angular momentum to be partly or fully
centrifugally supported reaches small
radii~(cf.~\cite{sekiguchi:10c}).  Based on progenitor structure, our
choice of rotation law, and the assumption of near free fall, we
estimate that this will occur after $\sim$$1.4\,\mathrm{s}$,
$\sim$$2.4\,\mathrm{s}$, $\sim$$3.9\,\mathrm{s}$ in model u75rot2,
u75rot1.5, u75rot1, respectively. At these times, the BHs, in the same
order, will have a mass ($a^\star$) of $\sim$$8\,M_\odot$ ($0.75$),
$\sim$$14\,M_\odot$ ($0.73$), and $\sim$$23\,M_\odot$ ($0.62$).

\begin{figure}
  \vspace{-0.5cm}
  \includegraphics[width=1.\linewidth]{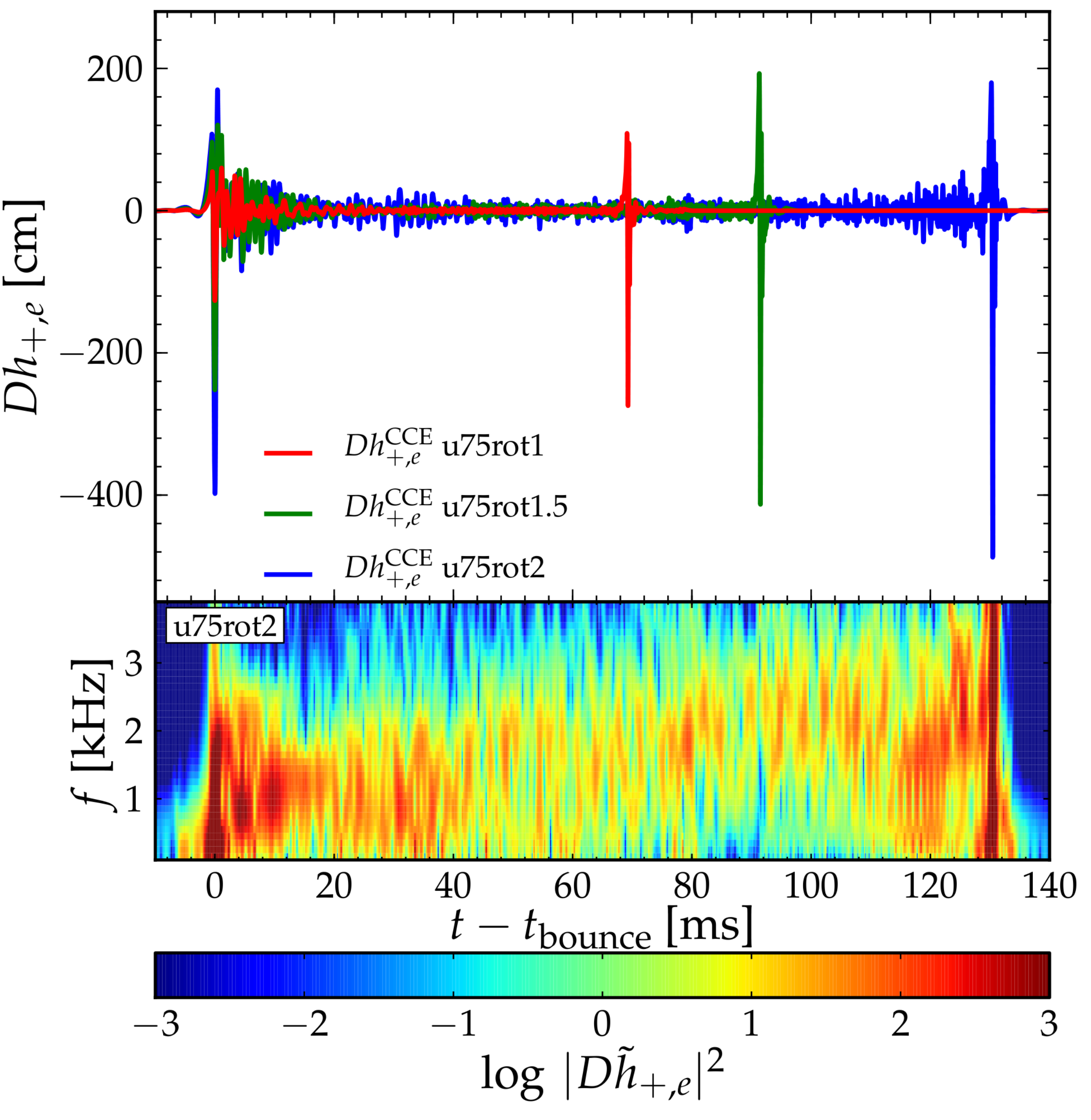}
  \vspace{-0.8cm}
  \caption{{Top}: GW signals $h_{+,e}$ emitted by the rotating models
  as seen by an equatorial observer and rescaled by observer distance
  $D$. {Bottom:} Spectrogram of the GW signal emitted by the most
  rapidly spinning model u75rot2.}
  \label{fig:gw}
\end{figure}

\emph{GW Signature.}---%
The top panel of Fig.~\ref{fig:gw} depicts the GW signals emitted by
our rotating models. Due to the assumed octant symmetry, GW emission
occurs in the $l=2, m=0$ mode.  The nonrotating model leads to a
very weak GW signal and is excluded. At bounce, a strong burst of GWs
is emitted with the typical signal morphology of rotating core
collapse~(e.g., \cite{ott:09}) and the peak amplitude is roughly
proportional to model spin. Once the bounce burst has ebbed, the
signal is dominated by emission from turbulence behind the shock. 
It is driven first by the negative entropy gradient left by the stalling
shock and then by neutrino cooling, whose effect may be overestimated
by our simple treatment.
Interestingly, the signal
strength increases with spin. This is not expected in a rapidly
spinning ordinary 2D CCSN, since a positive $j$ gradient in the
extended
postshock region stabilizes
convection.  In our models, the postshock region is considerably
smaller and shrinks with postbounce time. The driving entropy
gradients are steeper and the change of $j$ in the postshock region is
smaller.  Also, in contrast to 2D, our 3D models allow high-mode
nonaxisymmetric circulation.  We surmise that the combination of these
features with increasing spin (feeding greater circulation) results
in a stronger GW signal.

The intermittent period of turbulent, low-amplitude GW emission ends
when PNS collapse sets in, leading to a second pronounced spike in the
waveform, marking BH formation. The collapse signal evolves into the
ringdown emission of the nascent BH that rapidly assumes Kerr shape.
The GW emission ceases soon after and the un-shocked axisymmetric
accretion flow does not excite, at appreciable amplitude, BH
quasi-normal modes that could emit GWs. The strength of the BH
formation signal scales with $\Omega_0$ and its $dE_\mathrm{GW}/df$
peaks at $\sim$3.9~kHz, $\sim$3.4~kHz, $\sim$2.9~kHz, in u75rot1,
u75rot1.5, and u75rot2, respectively.  The lower panel of
Fig.~\ref{fig:gw} shows the spectrogram of the GW signal in model
u75rot2.  There is a clear trend towards higher frequencies during the
postbounce pre-BH phase, but BH formation itself, while 
peaking in the kHz range, leads to significant emission also at lower
frequencies, which is favorable for detection by advanced
laser-interferometer GW observatories (aLIGOs).  In
Table~\ref{tab:models}, we provide quantitative results on the GW
emission in our model set. For an event at $10\,\mathrm{kpc}$, we
estimate optimal single-detector aLIGO signal-to-noise ratios
(see~\cite{thorne:87,reisswig:10ccwave})
of $\sim$36 (u75rot1), $\sim$68 (u75rot1.5), and $\sim$94 (u75rot2),
and $\sim$6 for the nonrotating model u75rot0. Note that real GW burst
searches will not recover all available signal power.


\emph{Discussion.}---
We have performed self-consistent $3+1$ GR simulations of stellar
collapse in the context of the collapsar scenario for LGRBs. Albeit
approximate in many aspects, our models elucidate characteristic
qualitative features in the dynamics and GW signature of these events.
The rotating-collapse--bounce--PNS-phase--BH-formation--hyperaccretion
sequence and its GW signature are robust aspects of the early
collapsar evolution. More realistic physics will undoubtedly affect
quantitative results, but the overall qualitative picture is unlikely
to change. The characteristic GW signature seen in our models will
enable aLIGO to distinguish between a successful and failed
CCSN purely on the basis of observed GWs, provided the event is
sufficiently nearby.

A more realistic, stiffer EOS will increase the delay between bounce
and BH formation and will lead to higher-amplitude, lower-frequency
GWs. An improved neutrino treatment may reduce the vigor of
turbulence in the PNS phase and decrease the amplitudes of the
associated GW signal. Symmetry-free 3D evolution could reveal
nonaxisymmetric dynamics that may lead to an enhanced GW
signal~\cite{ott:09}. Only the inclusion of MHD may lead to a large
qualitative change by potentially leading to a strong explosion,
leaving behind a
magnetar~\cite{burrows:07b,dessart:08a,bucciantini:09}.  This study is
a first pioneering step and much work lies ahead before a clear and
quantitative picture of the CCSN-LGRB connection can be drawn.


This work was supported in part by NSF under grant nos.~AST-0855535,
OCI-0905046, PIF-0904015, PHY-0960291, and TG-PHY100033 and by the
Sherman Fairchild Foundation.  The GW signal data are available for
download at {\tt http://www.stellarcollapse.org/gwcatalog}.


\end{document}